\documentclass[fleqn,usenatbib,useAMS]{mnras}
\pdfoutput=1		% Because pdf plots are much smaller than eps ones and arxiv likes small files

\usepackage{txfonts}
\usepackage[T1]{fontenc}
\usepackage{ae,aecompl}

\usepackage{graphicx}	% Including figure files
\usepackage{amsmath}	% Advanced maths commands
\usepackage{amssymb}	% Extra maths symbols

\title[Galaxy structures and inclination biases]{Revealing strong bias in common measures of galaxy properties using new inclination-independent structures}

\author[B. M. Devour \& E. F. Bell]{
Brian M. Devour,$^{1}$\thanks{E-mail: bdevour@umich.edu}
Eric F. Bell,$^{1}$\thanks{E-mail: ericbell@umich.edu}
\\
$^{1}$Department of Astronomy, University of Michigan, 1085 South University Ave., Ann Arbor, MI 48109, USA
}

\date{\vspace{-0.4cm} Accepted 2017 February 1. Received 2017 January 31; in original form 2017 January 6}

\pubyear{2017}

\begin{document}
\label{firstpage}
\pagerange{\pageref{firstpage}--\pageref{lastpage}}
\maketitle

% Abstract of the paper
\begin{abstract}
Accurate measurement of galaxy structures is a prerequisite for quantitative investigation of galaxy properties or evolution. Yet, the impact of galaxy inclination and dust on commonly used metrics of galaxy structure is poorly quantified. We use infrared data sets to select inclination-independent samples of disc and flattened elliptical galaxies. These samples show strong variation in S\'{e}rsic index, concentration, and half-light radii with inclination. We develop novel inclination-independent galaxy structures by collapsing the light distribution in the near-infrared on to the major axis, yielding inclination-independent `linear' measures of size and concentration. With these new metrics we select a sample of Milky Way analogue galaxies with similar stellar masses, star formation rates, sizes and concentrations. Optical luminosities, light distributions, and spectral properties are all found to vary strongly with inclination: When inclining to edge-on, $r$-band luminosities dim by $>$1 magnitude, sizes decrease by a factor of 2, `dust-corrected' estimates of star formation rate drop threefold, metallicities decrease by 0.1 dex, and edge-on galaxies are half as likely to be classified as star forming. These systematic effects should be accounted for in analyses of galaxy properties.
\end{abstract}

% Select between one and six entries from the list of approved keywords.
% Don't make up new ones.

%%%%%%%%%%%%%%%%%%%%%%%%%%%%%%%%%%%%%%%%%%%%%%%%%%

%%%%%%%%%%%%%%%%% BODY OF PAPER %%%%%%%%%%%%%%%%%%

\section{Introduction}

An accurate understanding of galaxy structure -- the physical, three-dimensional distribution of light and/or mass -- is a cornerstone of the study of galaxies. Apart from their purely descriptive value, the structures of galaxies give insight into a range of physical ingredients and processes. Galaxy sizes and brightness profiles encode angular momentum content and its evolution (e.g. \citealt{mao_mo_white_98}; \citealt{vandenbosch_etal_01}). Galactic structure correlates with the orbital structure of stars in  galaxies and thus captures the demographics of rotation-supported and dispersion-supported galaxies \citep{vanderwel_etal_09_b}. Galaxy structure correlates strongly with star formation history (e.g. \citealt{kauffmann_etal_03_b}; \citealt{franx_etal_08}). Redshift evolution of the mix of galaxy structures quantifies disc and spheroid growth \citep{vanderwel_etal_14}, changes in the merger rate \citep{jogee_etal_09}, and the emergence of quiescent galaxies (e.g. \citealt{bell_etal_12}; \citealt{lang_etal_14_b}). As such, the astronomical community has devised many types of structure measurements -- e.g., S\'{e}rsic index \citep{sersic_63}, concentration \citep{strauss_etal_02}, $f_{\mathrm{DeV}}$ \citep{abazajian_etal_04}, and bulge-to-disc or bulge-to-total ratios \citep{dejong_96_ii}.

Ideally, a structural measurement should provide an unbiased description of a galaxy's light distribution. Yet, the projection of the light distribution on to the plane of the sky is a central challenge. Inclination differences lead to the same galaxy being mapped on to different two-dimensional projections owing both to geometry and varying dust attenuation with inclination. Unless structural measurements are designed to be viewing angle-independent, this will lead to inclination dependence in these measurements. Radiative transfer models suggest that structural measures should be wavelength- and inclination-dependent (e.g. \citealt{mollenhoff_etal_06}; \citealt{pastrav_etal_13_b}), in accord with measurements of more concentrated light profiles at longer wavelengths for dusty galaxies (e.g. \citealt{vulcani_etal_14}). Unfortunately, the \emph{inclination}-driven biases of many commonly used structural metrics are currently poorly understood.

The goal of this \emph{Letter} is to quantify the inclination dependence of commonly-used structural measurements and illustrate the impact of these systematic effects on our understanding of galaxy properties. We use inclination-independent metrics to select samples of similar galaxies, revealing that common structural measurements systematically vary with inclination for a variety of galaxy types (\S2). In \S3 we introduce a novel \emph{inclination-independent} technique for measuring galaxy structures. In \S4 we use these new measures to select an inclination-unbiased sample of Milky Way analogues and quantify the inclination dependence of their SDSS optical catalog quantities. All magnitudes are in the AB system, all logarithms are base-10, and where necessary we assume $\Omega_{\mathrm{M}} = 0.3$, $\Omega_{\Lambda} = 0.7$, and $H_{0} = 70 \ \mathrm{km} \ \mathrm{s}^{-1} \ \mathrm{Mpc}^{-1}$. \vspace{-0.4cm}

\section{Widely used structural metrics are biased by inclination}

Our analysis technique is conceptually simple. Using a sample of intrinsically similar galaxies observed from a variety of viewing angles, one can measure how galactic properties vary with inclination. Since the galaxies are intrinsically similar, differences between edge-on and face-on measures represent systematic errors due to dust and geometry. This method is used widely to measure the attenuation of galaxy luminosity with inclination (e.g. \citealt{tully_etal_98}; \citealt{devour_bell_16}); our use of it to explore the inclination dependence of galaxy structures and other properties is relatively novel.

This method requires that the metrics used to select the `intrinsically similar' galaxies do not depend on inclination. If an inclination-dependent metric (e.g., attenuation-dependent optical luminosity) is used to select similar galaxies, then the low and high inclination members of that sample are substantially different galaxies. Therefore, the critical challenge is to devise selection metrics which are inclination-independent. We briefly review such a selection and its properties here; for full discussion see \citet{devour_bell_16}.

Our initial galaxy sample consists of the cross-match of all galaxies with both elliptical aperture photometry from the \emph{Wide-field Infrared Survey Explorer} (\emph{WISE}; \citealt{wright_etal_10}) and spectroscopic redshifts from the Sloan Digital Sky Survey (SDSS) DR10 (\citealt{eisenstein_etal_11}, \citealt{ahn_etal_14}). We then select galaxies based on their \emph{WISE} elliptical aperture W1 (3.4\micron; henceforth $M_{3.4\micron}$) absolute magnitudes and \emph{WISE} W3--W1 ($12\micron - 3.4\micron$) colours (henceforth [12]--[3.4]), as illustrated in Fig.\ 1. Dust attenuation at these wavelengths is small, $<0.1$ mag attenuation between edge-on and face-on \citep{devour_bell_16}. Dust emission at $3.4\micron$ is also small, and variations in stellar M/L are expected to be no more than 0.3 dex \citep{meidt_etal_14}, so $M_{3.4\micron}$ luminosity is a reasonable proxy for stellar mass. Meanwhile, $12\micron$ emission arises from warm dust and PAH molecules \citep{calzetti_13}, and for star forming galaxies tracks overall star formation rate (SFR) to within $\sim$0.1 dex \citep{wen_etal_14}, making [12]--[3.4] colour a good proxy for SFR per unit stellar mass (specific SFR).

We adopt the $r$-band disc inclinations from the two-component S\'{e}rsic models of \citet{simard_etal_11} to derive axis ratios $a/b$. For intrinsically flattened galaxies (like the ones examined here) the axis ratio is a direct proxy for inclination, particularly as these model fit axial ratios account for the presence of a bulge in disc galaxies (which would otherwise increase the measured axis ratio). Ideally one would use a near-infrared axis ratio measurement to avoid any possible dust effects; in practice the axis ratios of \citet{simard_etal_11} are the best currently available. (For full discussion of the suitability of this metric, see \citealt{devour_bell_16}.)

\begin{figure}
\begin{center}
\includegraphics{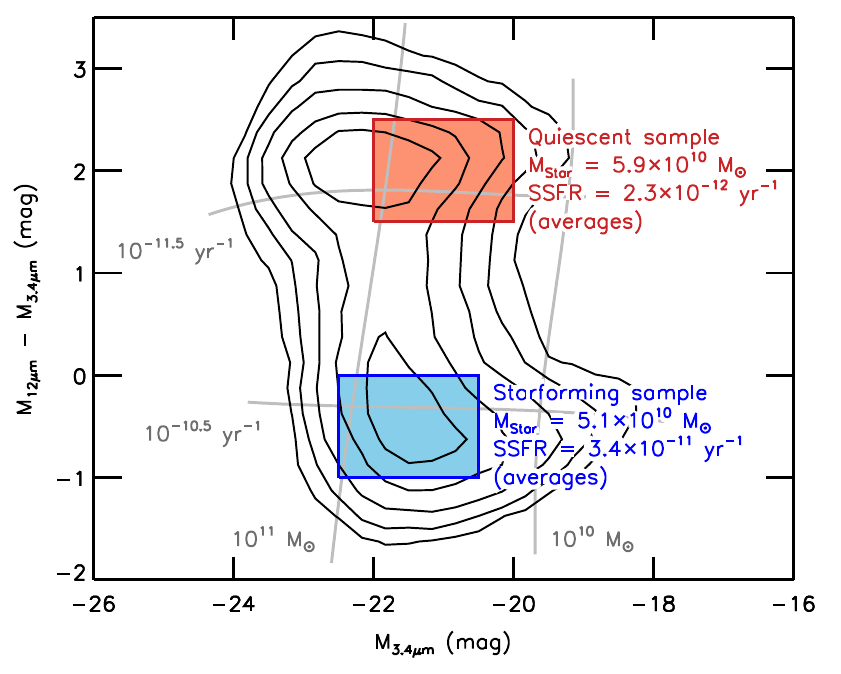}
\caption{Contours show the distribution of galaxies in $M_{3.4\micron}$ - [12]--[3.4]. Grey lines and labels show approximate values of stellar mass and specific SFR. Colour insets show the locations of our quiescent and star-forming samples and their average properties.}
\end{center}
\end{figure}

\begin{figure*}
\begin{center}
\includegraphics{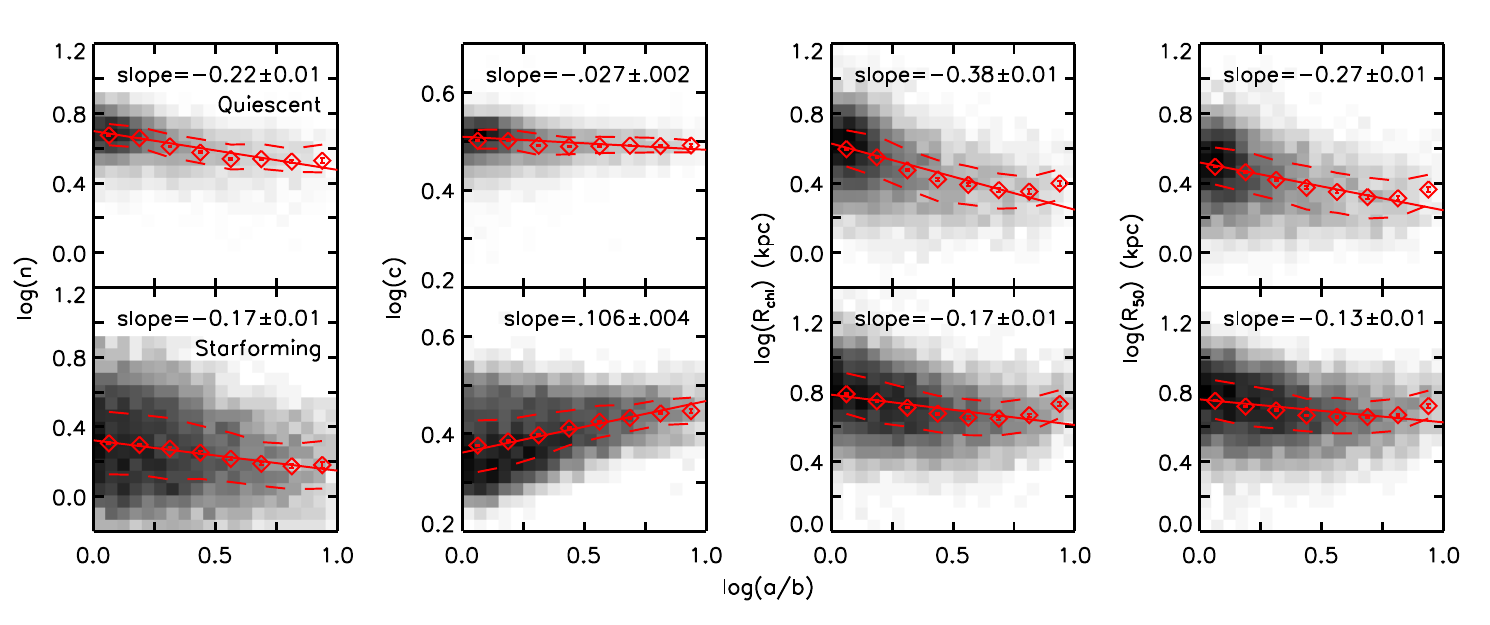}
\caption{Trends in structural parameters and sizes with $\log_{10}(a/b)$ for quiescent (top) and star-forming (bottom) disc galaxies. From left to right: S\'{e}rsic index $n$, circular concentration $c$, S\'{e}rsic fit circular half-light radius $R_{\rm chl}$, and circular Petrosian half-light radius $R_{50}$. Diamonds and dashed lines show the sample median and interquartile range, and the solid line shows a linear fit to the running median.}
\end{center}
\end{figure*}

By selecting galaxies as above (Fig.\ 1), we quantify how galaxy structure measurements vary with inclination for samples of intrinsically similar disc and flattened elliptical galaxies. We select star forming galaxies to have $-22.5 \leq M_{3.4\micron} \leq -20.5$ and $-1.0 \leq [12]-[3.4] \leq 0$, corresponding to star forming main sequence galaxies with  $3\times10^{10}<M_*<6\times10^{10} \ \mathrm{M}_{\sun}$. Our quiescent sample has $-22.0 \leq M_{3.4\micron} \leq -20.0$ and $1.5 \leq [12]-[3.4] \leq 2.5$, corresponding to discy quiescent galaxies with $4\times10^{10}<M_*<8\times10^{10} \ \mathrm{M}_{\sun}$. The parent sample contains 78,721 galaxies; these subsamples contain 7044 and 6239 galaxies, respectively, 75 per cent of which have $0.0279 \leq z \leq 0.0635$. 

In Fig.\ 2, we show the variation with inclination (quantified by $\log_{10}(a/b)$) of $r$-band S\'{e}rsic indices $n$ and circular half-light radii $R_{\rm chl}$ from the two-component S\'{e}rsic fits of \citet{simard_etal_11}, and $r$-band concentrations $c$ and Petrosian circular half-light radii $R_{50}$ from SDSS. In each panel, we show the slope of a linear fit to the running median of that structural parameter with $\log_{10}(a/b)$; uncertainties in the slopes are calculated using bootstrap resampling.

Both S\'{e}rsic index and concentration strongly depend on inclination for star-forming disc galaxies, likely reflecting some combination of differential dust attenuation and geometric effects. Interestingly, S\'{e}rsic index also shows a strong dependence on inclination for the flattened quiescent (dust-free) galaxies. Concentration shows almost no trend with axis ratio for quiescent galaxies. Both S\'{e}rsic-derived half-light radius $R_{\rm chl}$ and Petrosian half-light radius $R_{50}$ vary significantly with inclination, decreasing significantly and then increasing slightly in size as galaxies tilt towards edge-on.

This is a major concern; commonly used metrics of galaxy structure depend on inclination at a level comparable to the dispersion of structural metrics within a population.
\vspace{-0.4cm}

\section{inclination-independent galaxy structural measurements}

There are two primary obstacles to  inclination-independent measurements of galaxy structure. 

The first is dust attenuation. Currently it is impossible to dust-correct the optical light distribution to the accuracy demanded by current analyses. Accordingly, we choose to analyse the galaxy structures in $K$ band (2.2\micron) from the UKIRT Infrared Deep Sky Survey (UKIDSS) Large Area Survey \citep{lawrence_etal_07}. Dust attenuation at $K$ band is $1/5-1/3$ of the optical attenuation (depending on poorly understood radiative transfer effects; \citealt{tuffs_etal_04}), and is generally $<0.3$ mag even for nearly edge-on galaxies with high optical opacities; galaxies are therefore close to transparent across almost all viewing angles in $K$ band, avoiding significant dust-induced inclination dependence.

\begin{figure*}
\begin{center}
\includegraphics{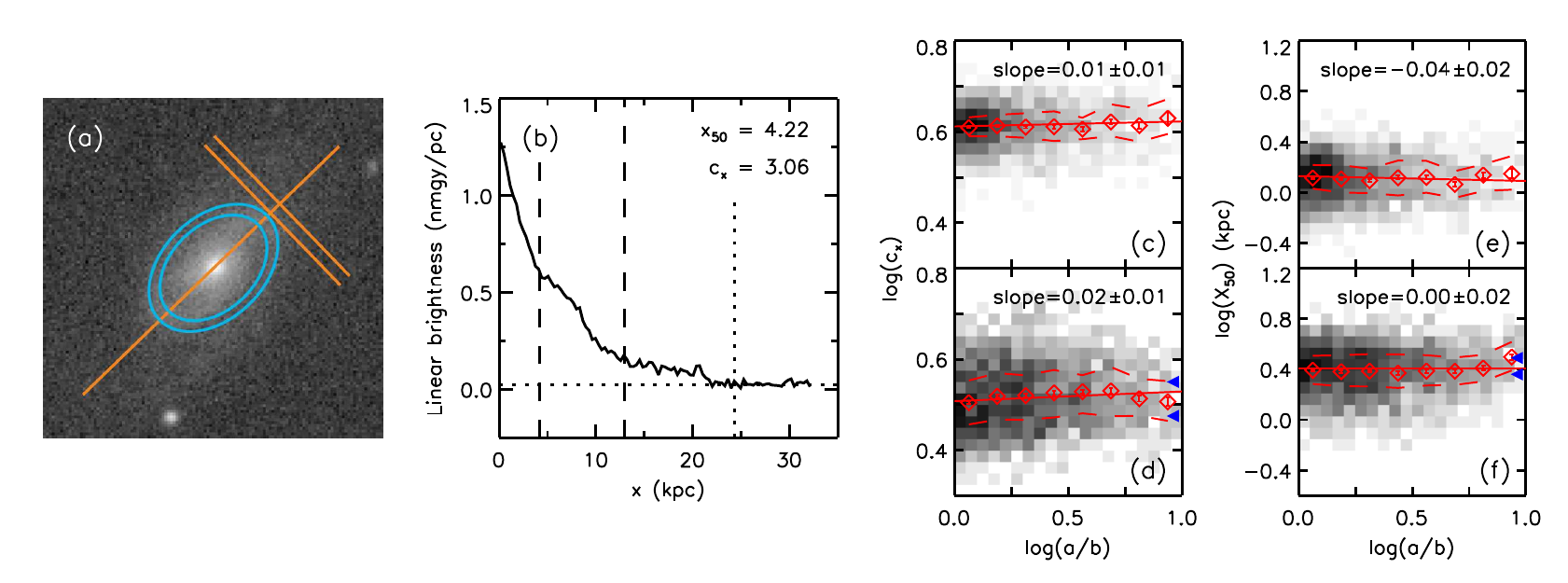}
\caption{Panel a: $K$-band galaxy stamp, with example elliptical annulus and major axis cut. Panel b: Folded linear brightness profile and structural measurements. Dotted lines indicate the measured sky level and profile `edge' location, and dashed lines indicate measured $x_{50}$ and $x_{90}$. Panels c-f: Trends in linear concentration $c_{\rm x}$ (left) and linear half-light distance $x_{50}$ (right) with $\log_{10}(a/b)$ for quiescent (top, panels c and e) and star-forming (bottom, panels d and f) disc galaxies. Diamonds and dashed lines show the median and interquartile range, and the solid line shows a linear fit to the running median. Blue carets in panels d and f bracket the $c_{\rm x}$ and $x_{50}$ selections in \S4.}
\end{center}
\end{figure*}

The second obstacle is the differing projection suffered by face-on and edge-on galaxies. Many structural parameters are calculated using elliptical or circular apertures or annuli, as illustrated in Fig.\ 3 (blue rings in panel a). Circular annuli and apertures are obviously strongly affected by inclination, as the portions of a galaxy that are projected within a circular aperture change dramatically with inclination. Elliptical annuli/apertures mitigate this obvious issue, but still suffer from inclination-dependent projection effects. An elliptical annulus only maps to a region of constant galactocentric radius in the limit of an infinitely thin disc. Towards larger inclination, an elliptical annulus becomes substantially more influenced by the vertical structure of a disc and mixes light from a variety of galactocentric radii (e.g. for a disc with an intrinsic $b/a=0.1$ viewed from 60$^{\circ}$\ inclination, an elliptical annulus mixes light from a range of radii corresponding to $\sim$35 per cent of its radial scale length). Furthermore, any variations in the vertical light profiles of galaxies will lead to further variation of structural measurements derived using elliptical annuli or apertures.

The issues with traditional metrics arise from the varying projection of light along the minor axis with inclination. Our technique sidesteps this issue. We take cuts parallel to the minor axis, collapsing all light down on to the major axis to form a \emph{linear}, rather than radial, brightness profile, as illustrated in Fig.\ 3 (orange lines in panel a). This method has the decisive advantage that in the absence of dust attenuation it is inclination-independent \emph{by construction}.

Each galaxy's linear brightness profile is folded across the centre, correcting for contamination and sky background. We then measure the 50 and 90 per cent distances (henceforth $x_{50}$ and $x_{90}$) by analogy with $r_{50}$ or $r_{90}$ in circular or elliptical aperture photometry -- these are the linear distances from the centre that contain 50 or 90 per cent of the total brightness. The linear concentration (henceforth $c_{\rm x}$) is then the ratio $x_{90}/x_{50}$, again in analogy to the traditional measurement. We illustrate this process in panel b of Fig.\ 3. The linear sizes $x_{50}$ and concentrations $c_{\rm x}$ are then inclination-independent structural parameters. A full discussion of the process, uncertainties, and the galaxy parameter distribution will be presented in a forthcoming paper.

These parameters show little systematic dependence on inclination for either disc-dominated star forming or quiescent galaxies (Fig.\ 3, rightmost panels). A linear fit to the running median of log($c_{\rm x}$) or log($x_{50}$) (kpc) with log($a/b$) gives variations between face-on and edge-on of $<0.04$ dex (c.f., variations of 0.1--0.4 dex for most widely-used galaxy structures in Fig.\ 2). The distributions of $c_{\rm x}$ and $x_{50}$ appear noisier owing to smaller sample sizes, as only $\sim$30 per cent of our sample has UKIDSS imaging, but the \emph{scatter} in $c_{\rm x}$ or $x_{50}$ is comparable to the scatter in S\'{e}rsic index and concentration, or $R_{\rm chl}$ and $R_{50}$, respectively, for the same populations. Typical uncertainties in $c_{\rm x}$ and $x_{50}$ are $<0.1$--0.2 dex respectively, thus, much of the scatter in $c_{\rm x}$ and $x_{50}$ in a given bin of \emph{WISE} absolute magnitude and colour is intrinsic.
\vspace{-0.4cm}

\section{How does our view of Milky Way analogues vary with inclination?}

\begin{figure}
\begin{center}
\includegraphics{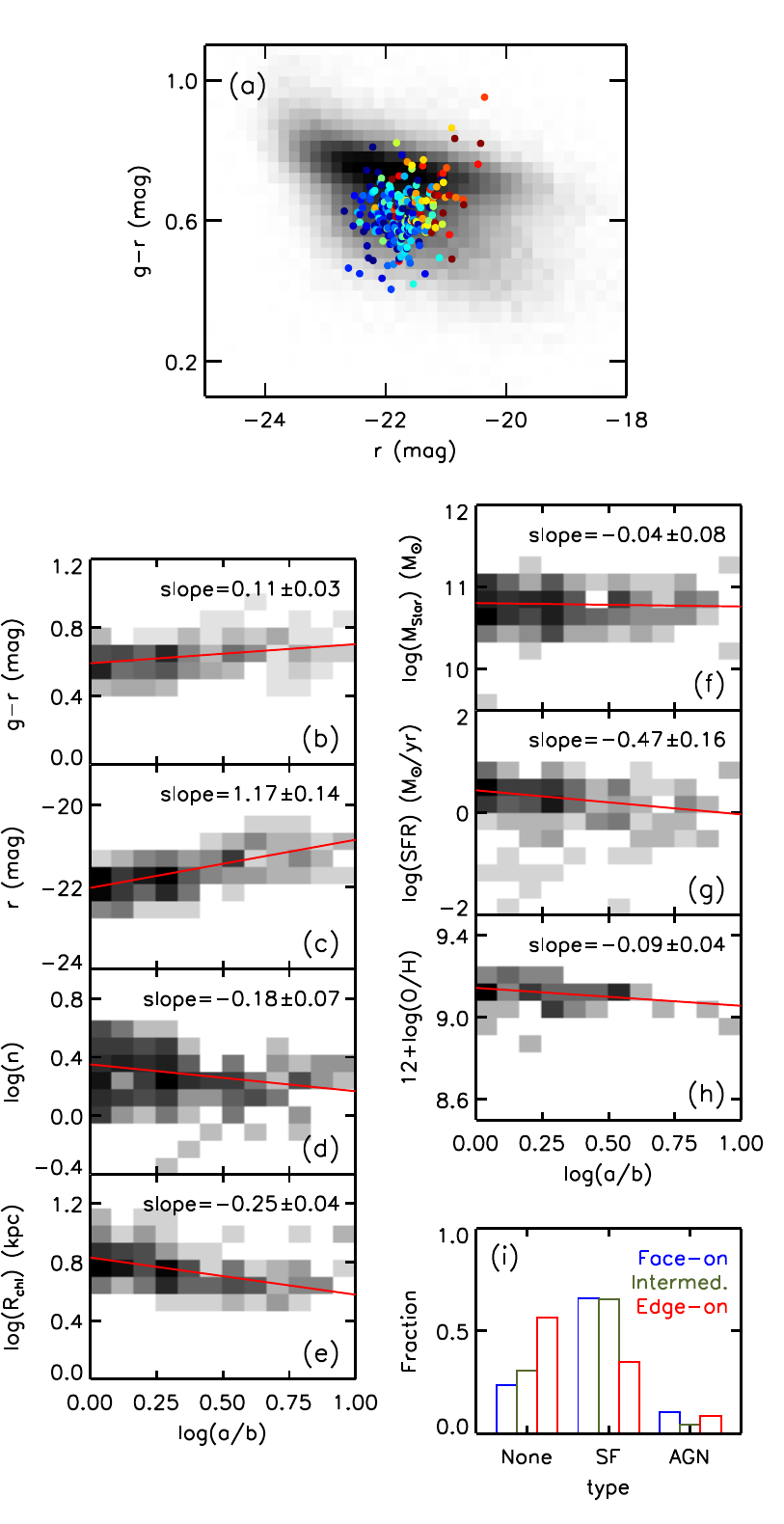}
\caption{Panel a: Colour-magnitude diagram for the parent sample (grey histogram) and our sample of star-forming Milky Way analogue disc galaxies (overplotted points). Bluer colours show face-on galaxies and redder colours show edge-on galaxies. Panels b-e: Top to bottom, the distribution of our sample's $g$--$r$ colours, $r$-band absolute magnitudes, S\'{e}rsic indices $n$, and circular half-light radii $R_{\rm chl}$ as a function of $\log(a/b)$. Panels f-i: Top to bottom, the distribution of our sample's SDSS estimates of stellar mass, optically-derived star formation rate, and metallicity as a function of $\log(a/b)$, and distribution of spectral classification for subsamples with low, moderate, and high inclination.}
\end{center}
\end{figure}

With these inclination-independent structural metrics, we can sharpen our selection of intermediate-mass star-forming main sequence disc galaxies to also have a relatively narrow range in linear concentrations and sizes: We select galaxies to have $-22.25 \leq M_{3.4\micron} \leq -20.75$, $-1.0 \leq [12]-[3.4] \leq 0$, $0.475 \leq \log(c_{\rm x}) \leq 0.55$, and $0.36 \leq \log(x_{50}) \ (\mathrm{kpc}) \leq 0.49$. (The $c_{\rm x}$ and $x_{50}$ limits are indicated in Fig.\ 3, and this selection is similar in [12]--[3.4] and slightly narrower in $M_{3.4\micron}$ than the star-forming sample depicted in Fig.\ 1.) The Milky Way's stellar mass and specific SFR have been estimated as roughly $M_{*} = 5 \times 10^{10} \ \mathrm{M}_{\sun}$ and $\mathrm{SSFR} = 3 \times 10^{-11} \ \textrm{yr}^{-1}$ \citep{bland-hawthorn_gerhard_16}, placing it within this bin (see Fig. 1). Thus, this sample consists of structurally average star-forming disc galaxies broadly similar in stellar mass and SFR to the Milky Way. Following \citet{licquia_etal_15}, we refer to these as `Milky Way analogues', but with the important advantages that this sample is also known to be structurally typical and is selected in a way that is insensitive to inclination and dust attenuation. With it, we can quantify how commonly used optical SDSS metrics respond to inclination and dust for a sample of galaxies that are known to be intrinsically very similar to each other.

Fig.\ 4 shows our main results. In panel a, we see that the face-on members of our sample (bluer points) clearly fall within the star-forming `blue cloud'. The more highly inclined members (redder points) become both redder and fainter, in some cases overlapping with the `red sequence' of galaxies usually interpreted as quiescent. This variation in absolute magnitude and colour with axis ratio is quantified in panels b and c. Panels d and e confirm that SDSS optical structural measurements strongly depend on axis ratio, as foreshadowed in Figs.\ 1 and 2, and now verified and sharpened for a sample selected to have the same intrinsic light profiles. In panels f--i we show optical SDSS spectral line-derived galaxy properties: stellar mass, dust-corrected SFR, and metallicity from \citet{kauffmann_etal_03}, \citet{tremonti_etal_04}, and \citet{brinchmann_etal_04}, and spectral classification from \citet{bolton_etal_12}. Stellar mass is not significantly affected by dust, as expected (\citealt{bell_dejong_01}; \citealt{maller_etal_09}). In contrast, spectral line-based SFRs -- meant to include dust corrections -- emission-line derived metallicities, and spectral classifications all clearly vary with inclination for this sample of intrinsically similar galaxies. Optically-derived and `dust-corrected' SFRs and spectral classifications are particularly strongly affected: SFRs for edge-on members of this sample appear to be only $1/3$ of the SFRs of their face-on intrinsically similar counterparts, and edge-ons are only half as likely to be classified as star forming according to their emission line diagnostics. 

The sample explored in Fig.\ 4 is selected to consist of intrinsically similar galaxies, with similar stellar masses, SFRs, sizes and concentrations, save for their viewing angles. And yet, \emph{optical SDSS measurements of galaxy properties would conclude that the more edge-on members of this population are systematically dimmer, redder, less centrally concentrated, smaller, less actively star-forming, and lower metallicity than their face-on counterparts}. Optical observations of galaxy luminosities, light distributions, and spectral lines are all affected; it is clear that one cannot isolate samples of intrinsically similar galaxies for study without inclination-independent and IR-based selection tools.
\vspace{-0.4cm}

\section{Conclusions and Outlook}

In this \emph{Letter}, we use inclination-independent infrared metrics to select samples of intrinsically similar star forming disc galaxies and quiescent galaxies. We show that traditional structural measurements suffer from severe inclination biases, showing variations from edge-on to face-on comparable to the spread of parameters within each population.

We present an alternative method for measuring structural parameters (the linear brightness profile) which is inclination independent by construction. When used with dust-penetrated NIR imagery, this technique gives inclination-independent measures of galaxy size and structure. 

We then use our inclination-independent techniques to select a sample of star-forming disc galaxies with very similar intrinsic masses, SFRs, sizes and concentrations that are expected to be broadly similar to the Milky Way. While SDSS optical estimates of stellar mass are robust to inclination and dust effects, \emph{measures of optical luminosity, colour, light profile shape, galaxy size, spectral classification, optically-derived `dust-corrected' SFR and emission line metallicity all depend strongly on inclination and dust attenuation.}  

These systematic errors in SDSS structural and spectral measurements may impact a wide range of studies. Samples selected by SFR or colour -- green valley galaxies, red discs, starbursts -- may all be affected by dust in an inclination-dependent fashion. The census and properties of bulges in disc-dominated galaxies will be affected. Our work highlights the importance of IR imagery in understanding these effects, an important consideration given the upcoming launch of the \emph{James Webb Space Telescope}. We will explore some of these issues in future papers, and release a dust-independent size and concentration catalogue to allow others to quantify the effects of inclination and dust on their analyses.

\section*{Acknowledgements}

This work was supported by NSF-AST 1514835. We use data products from the \emph{Wide-field Infrared Survey Explorer}, which is a joint project of the University of California, Los Angeles, and the Jet Propulsion Laboratory/California Institute of Technology, funded by the National Aeronautics and Space Administration. Funding for the Sloan Digital Sky Survey IV has been provided by the Alfred P. Sloan Foundation, the U.S. Department of Energy Office of Science, and the Participating Institutions. The SDSS web site is www.sdss.org. When the data reported here were acquired, UKIRT was operated by the Joint Astronomy Centre on behalf of the U.K. Science and Technology Facilities Council.
\vspace{-0.8cm}

% The first author would also like to thank Panda Express for calories, Espresso Royale for caffeine, and the residents of the New Eden cluster for many hours of delightful explodey distraction.

%%%%%%%%%%%%%%%%%%%%%%%%%%%%%%%%%%%%%%%%%%%%%%%%%%

%%%%%%%%%%%%%%%%%%%% REFERENCES %%%%%%%%%%%%%%%%%%

% The best way to enter references is to use BibTeX:

\bibliographystyle{mnras}
\bibliography{combinedbib}

%%%%%%%%%%%%%%%%%%%%%%%%%%%%%%%%%%%%%%%%%%%%%%%%%%

% Don't change these lines
\bsp	% typesetting comment
\label{lastpage}
\end{document}